\documentclass[12pt]{article}
\usepackage{fullpage,authblk,cite}
\usepackage{amsmath,amsthm,graphicx,amssymb,verbatim,listings}
\usepackage{wasysym,color,enumitem,mathcomp}
\usepackage[T1]{fontenc}     
\usepackage{lmodern, tipa}   
\usepackage[ pdftex, plainpages = false, pdfpagelabels,
                 pdfpagelayout = useoutlines,
                 bookmarks,
                 bookmarksopen = true,
                 bookmarksnumbered = true,
                 breaklinks = true,
                 linktocpage=all,
                 pagebackref=false,
                 colorlinks = true,
                 linkcolor = Blue,
                 urlcolor  = blue,
                 citecolor = BrickRed,
                 anchorcolor = green,
                 hyperindex = true,
                 hyperfigures
                 ]{hyperref}
\usepackage[usenames, dvipsnames]{xcolor}
\usepackage{xifthen} 

\newcommand{\tr}{\hbox{tr}}

\newcommand{\ket}[1]{{\ensuremath{\left| #1 \right\rangle}}}
\newcommand{\bra}[1]{{\ensuremath{\left\langle #1 \right|}}}

\newcommand{\ketbra}[2]{{\ensuremath{\left| #1 \middle\rangle \middle\langle #2
      \right|}}}

\newcommand{\arxiv}[2][]{\ifthenelse{\isempty{#1}}{\href{http://arxiv.org/abs/#2}{{\tt arXiv:\allowbreak{}#2}}} {\href{http://arxiv.org/abs/#2}{{\tt arXiv:\allowbreak{}#2 [#1]}}}}

\newcommand{\booktitle}{\textsl}
\newcommand{\hrefdoi}[2]{\href{https://dx.doi.org/#1}{#2}}

\newcommand{\bbC}{\mathbb{C}}

\begin{document}
\title{Masanes--Galley--M\"uller and the State-Update Postulate}
\author[$\dag$]{Blake C.\ Stacey}
\affil[$\dag$]{Physics Department, University of
    Massachusetts Boston\protect\\ 100 Morrissey Boulevard, Boston MA 02125, USA}

\date{\small\today}

\maketitle

\begin{abstract}
  Masanes, Galley and M\"uller claim to have derived a unique rule for
  quantum state update consequent upon a measurement outcome. Upon
  closer examination, their proof implicitly assumes its first step,
  namely that the state-update rule is linear.
\end{abstract}

In this note, I am discussing ``The measurement postulates of
quantum mechanics are operationally redundant'' by Masanes, Galley and
M\"uller~\cite{Masanes:2019}. MGM begin by listing what they call
the five ``standard postulates'' of quantum
theory. Then, they replace the last two with axioms of a more
qualitative nature, and they re-derive the standard set. The two
postulates they wish to replace are, first, the Born rule for POVM
elements applied to pure states:
\begin{quote}
  Each measurement outcome of [a system with Hilbert space] $\bbC^d$
  is represented by a linear operator $Q$ on $\bbC^d$ satisfying $0
  \leq Q \leq I$, where $I$ is the identity. The probability of
  outcome $Q$ on state $\psi \in \bbC^d$ is
  \begin{equation}
    P(Q|\psi) = \bra{\psi}Q\ket{\psi}.
  \end{equation}
\end{quote}
They go on to give the familiar definition of a POVM, saying that a
full set of measurement outcomes corresponds to a set of linear
operators that add to the identity. This is unremarkable. The second
postulate they wish to replace is that of ``post-measurement state
update'':
\begin{quote}
  Each outcome is represented by a completely-positive linear map
  $\Lambda$ related to the operator $Q$ via
  \begin{equation}
    \tr \Lambda(\ketbra{\psi}{\psi}) = \bra{\psi} Q \ket{\psi},
  \end{equation}
  for all $\psi$. The post-measurement state after outcome $\Lambda$
  is
  \begin{equation}
    \rho = \frac{\Lambda(\ketbra{\psi}{\psi})}
         {\tr \Lambda(\ketbra{\psi}{\psi})}.
         \label{eq:huh}
  \end{equation}
\end{quote}

MGM's justification for their claim is in Supplementary note 5, where
it is designated Lemma 22.
\begin{quote}
  At this point we still have not said anything about the
  post-measurement state update rule. But whatever this rule is, let
  $\sigma(F_i, \rho)$ be the post-measurement state (that is, its
  density matrix) after outcome $F_i$, when the initial state is
  $\rho$. And define the map $\Lambda_{F_i}$ which takes the original
  state $\rho$ to the post-measurement state times its corresponding
  probability:
  \begin{equation}
    \Lambda_{F_i}(\rho) := \sigma(F_i, \rho) \tr[F_i \rho].
  \end{equation}
\end{quote}
So far, this is just notation, and seems unobjectionable.
\begin{quote}
Next, consider another given measurement with POVM $\{G_j\}$, and
define the POVM $\{H_{j,i}\}$ to be that corresponding to the
successive implementation of the measurements $\{F_i\}$ and
$\{G_j\}$. (This must correspond to a valid measurement, because the
whole point of talking about a post-measurement state is that one can
make further measurements on it.) Then, using the rules of probability
calculus and the above formulas we obtain
\begin{align}
  \tr[H_{j,i} \rho] &= P(j,i) = P(j|i) P(i)
  = \tr[G_j \sigma(F_i,\rho)] \tr[F_i \rho] \nonumber \\
  &= \tr[G_j \Lambda_{F_i}(\rho)],
\end{align}
for all $i$, $j$ and $\rho$. This equation implies that the map
$\Lambda_{F_i}(\rho)$ is linear in $\rho$.
\end{quote}
In other words, the linearity of the map $\Lambda_{F_i}(\rho)$ follows
from the assumption that $H_{j,i}$ is a POVM element that can be used
as any other. But $\{H_{j,i}\}$ is not a measurement that is performed
at a single time. It is, by necessity, a sequence of two experimental
interventions, with a state-change in between. Indeed, ``the whole
point of talking about a post-measurement state is that one can make
further measurements on it,'' but the assumption that two-stage
experiments have their mathematical representations in exactly the
same set as one-stage experiments is an assumption that state-change
is undramatic. (We have already accustomed ourselves to the fact that
the \textsc{and} operation is not always meaningful for quantum
experiments; on what deep basis do we say it should apply trivially
here?) We could imagine, for example, that a measurement upon a qubit
yields results with Born-rule probabilities, and then transforms the
Bloch-sphere coordinates nonlinearly. We could have a first
measurement $\{F_i\}$ and a second measurement $\{G_j\}$, but the
transformation taking place in between would be nonlinear, and so the
conclusion about the map $\Lambda_{F_i}$ would simply not follow. For
instance, suppose the initial state $\rho$ is a mixture of the Pauli
eigenstate $\ketbra{z+}{z+}$ and the garbage state $\frac{1}{2}I$, and
let the state-update map implement a logistic transformation of the
Bloch-sphere radial coordinate:
\begin{equation}
  r \to r' = \lambda r(1-r),
\end{equation}
with $\lambda \in [0,4]$. If the first and second POVMs are both the
computational basis measurement $\{\ketbra{z+}{z+},
\ketbra{z-}{z-}\}$, then the probability of either outcome in the
second measurement is trivially a nonlinear function of the initial
state $\rho$. MGM write that ``it is well-known\textsuperscript{36}
that the only possible state-update rule that is compatible with'' the
Born rule is Eq.~(\ref{eq:huh}). Their reference 36 is to a 1970 paper
by Davies and Lewis~\cite{Davies:1970}, which assumes linearity of
state transformations.

In other words, assuming that an $\{H_{j,i}\}$ can be defined as in
MGM's Lemma 22 amounts to assuming the linearity of the state-update
map. It has been known, probably since Holevo in the
1970s~\cite[p.\ 973]{Fuchs:2014}, that a function $f$ which maps the
space of density matrices to the unit interval can be written as
$\tr(\rho E)$ for some effect operator $E$ if it is convex-linear:
\begin{equation}
  f\left(\sum_i p_i \rho_i\right) = \sum_i p_i f(\rho_i),
  \ \hbox{with}\ p_i \geq 0,\ \sum_i p_i = 1.
\end{equation}
The existence of the $H_{j,i}$ on which MGM's lemma depends follows
from the linearity they wish to prove. This linearity can be derived
from other premises, such as assumptions about
context-independence~\cite{Flatt:2017} as in the proofs of the
POVM-Gleason theorem~\cite{Busch:2003, Caves:2004}. However, these
assumptions do not appear to be entailed in the postulates that MGM
put forth. Indeed, MGM write, ``We stress that our results, unlike
previous contributions [\,\ldots], do not assume'' such premises.

\bigskip

I thank Chris Fuchs, R\"udiger Schack and Matt Weiss for discussions.

\bigskip

\textbf{Update (19 February 2023):} Masanes, Galley and M\"uller have
replied to this comment. I encourage the reader to examine their
\arxiv{2212.03629} before continuing, if they have not already done
so. It is, I believe, a clarifying statement.

They write, ``Our derivation of the Born rule is completely agnostic
as to how the outcome comes about or how many steps it takes.'' If
this had been stated in their original paper, there would be much less
to say now. Or, to put it differently, we would be contemplating a
wider variety of possible state-update rules as a natural weakening of
their assumptions, rather than an unwanted consequence of the
assumptions as written. As their response notes, their theorem is
inapplicable to situations like a system ``constrained by a
fundamental superselection rule, forbidding superpositions between
some charge sectors'', or to scenarios like Wigner's Friend that
involve ``a sequence of outcomes that, intuitively, do not necessarily
`co-exist'\,''. I concur that in these scenarios, ``it is unclear
whether, and how, it makes sense to assign (joint) probabilities to
these sequences of outcomes''. But the very issue that Wigner's Friend
is designed to raise is the question of what counts as a measurement
and what changes when one takes place. Why is it that Wigner assigning
joint probabilities over a sequence is dubious when he measures his
friend, but not when he measures an electron? Is the friend a
qualitatively different type of physical system?

The 2019 paper presents ``a theory-independent characterization of
measurements for single and multipartite systems''. This
characterization assumes no particular interpretation of probability
theory, but it does invoke the structure of rays in $d$-dimensional
complex Hilbert spaces, and tensor products of those spaces. It is
stated that ``the observer always has the option of describing a
system $\bbC^a$ as part of a larger system $\bbC^a \otimes
\bbC^b$''. What is not stated is that a sequence of events produced over
many steps, each of which is registered by the observer, can also be
regarded as a single ``measurement outcome''. The justification as
written in the Supplemental Material for why a two-stage measurement
can get the same mathematical treatment as a single-stage one is
\emph{not} agnosticism, but the claim that \emph{some} quantum state
must apply to the system in between stages. This is plainly weaker
than agnosticism. Must agnosticism be spelled out explicitly? I think
so. For one reason, doing so is a matter of living up to the same
standard that the paper does elsewhere. For another, agnosticism must
be delimited with care, because taking it too far leads one to make
statements about quantum mechanics that are actually false.

It is useful to compare the assumption of agnosticism with how MGM
define the completeness of the set of ``outcome probability
functions'' (OPFs). In the text leading up to Eq.~(7) of their
original paper, MGM spell out explicitly that the set of OPFs is
closed under convex combinations. This is a nontrivial assumption. If
I say that I can pick between two procedures with probability $p$ and
$1-p$ respectively, then either I have a source of randomness outside
the domain of my theory (like a coin with arbitrarily tunable bias),
or I am baking into my theory from an early stage a premise of what
kinds of probabilities it can output. Either way, if I make my
stochastic choice between procedures based on a datum from some
experiment, then the physical object that yielded that datum is in the
causal past of my lab equipment. What rules out an influence from my
coin to the rest of my apparatus? Why should doing a procedure with
that additional fact in hand imply the same statistics as doing it
without? (If Alice catches Bob flipping a coin to decide what he will
do in lab, she may justifiably conclude he is careless and
indifferent, thus changing her predictions about how well he will
handle the equipment, perhaps even concluding that some very expensive
machines are at risk.) It is not unwarranted to assume as a starting
point that convex combinations work out cleanly, or to find conceptual
grounds to justify that mathematical claim, but it \emph{is} an
assumption that needs being said before it can be invoked.

The same concerns apply to multi-stage operations. In both cases, life
is much simpler if we can map two-index objects into the same set as
one-index objects. Eq.~(7) of the 2019 paper \emph{needs} to be
included, and it is very good that the paper does. But if \emph{that}
statement about two-step procedures being mappable into the set of
one-step procedures deserves an explicit statement (``with circles and
arrows and a paragraph on the back'', as it were) then so does the
agnosticism assumption upon which the derivation of the state-update
rule relies.

Another concern regarding agnosticism can be illustrated by a
comparison with classical probability theory. Suppose that $X$ and $Y$
are two random variables. Then we can characterize them by a joint
probability distribution $P(X,Y)$ and take marginals by summation to
construct $P(X)$ and $P(Y)$. This works in the same way whether the
random variables $X$ and $Y$ stand for simultaneous phenomena, or
whether one is the future of the other. In either case, the object
from which marginals are taken satisfies the same mathematical
properties. The situation in quantum mechanics is more complicated. If
$X$ and $Y$ are two quantum systems at the same time, we assign them a
joint state $\rho_{XY}$ which is some positive semidefinite operator,
and we can extract the marginal states by partial traces. But if we
try to construct a ``joint state'' for the same system at two times,
an operator from which the present and future states are both partial
traces, it will in general not be positive
semidefinite~\cite{Leifer:2013, Horsman:2017}. The dual statement also
holds if we consider time evolution in the Heisenberg picture and
evolve our POVM elements instead of our states. Either way, we find
ourselves manipulating causal conditional operators that do not
generally belong to the same set as acausal operators
do~\cite{Leifer:2013}. Sometimes it matters how an outcome comes
about: the pair of indices $(i,j)$ needs to be handled differently
depending on the kind of causal relation between $i$ and $j$. If one
took the thoroughly agnostic position and assumed that pairs of events
should be treated in exactly the same way regardless of their temporal
ordering, then one would not get quantum theory. Indeed, depending on
how many of the properties of ordinary quantum states one wishes to
maintain for causal joint states, constructing a causal joint state
for three or more times can be
\emph{impossible}~\cite{Horsman:2017}. So, one really cannot afford to
be agnostic about how many events are in a sequence!

Agnosticism, carried too far, would lead us to conclude that it does
not matter, for example, whether a detector is placed at the slits in
a double-slit experiment. That too is a question of how an ``outcome
comes about or how many steps it takes''. No physicist would endorse
that position, but if we are to understand the logical structure of a
theory, we cannot rely on physicists' background knowledge to resolve
ambiguities. As the conditional-states formalism shows, we cannot even
make the weaker statement that the set of mathematical entities from
which we draw the description of measurement events should be
agnostic about how those events are arranged temporally.

MGM observe that the pathological model combining the Bloch ball with
the logistic map could be ruled out because it ``would allow for
discriminating two ensembles of the same mixed state''. Being a
deliberately bizarre chimera, it could doubtless be ruled out in many
ways! Rather than focusing on the Bloch sphere, we could instead
consider a model in which the statistics for single-step measurements
are encoded into density matrices via the Born rule, and upon
measurement the state updates by some CPTP map, the choice of which
depends not just upon the measurement outcome but also upon the
probability with which that outcome took place. Any model of that sort
will inevitably not play nicely with the existence of multiple convex
decompositions of the same density matrix. The people who take
nonlinear modifications of quantum mechanics seriously have noticed
this and concluded, sensibly enough, that scenarios which are
equivalent in quantum theory can be given inequivalent mathematical
representations in a not-quantum theory~\cite{Haag:1978,
  Jordan:1993}. For instance, a not-quantum theory might regard having
access to only part of a system as requiring a different class of
description than having a probabilistic choice among systems. In
quantum theory, we use a partial trace for the former and a mixture
for the latter, ending with a density matrix in either case, but not
every theory needs to have this property. This is another angle from
which to consider the point that a theory can make the same
predictions as quantum mechanics for single-shot experiments, but
differ in its predictions for multi-stage ones. However, inventing
various kinds of nonlinear dynamics is secondary to the topic of
\emph{kinematics,} i.e., whether having a measurement result in hand
is already a kind of ``superselection'' between sectors.

A measurement, in both ordinary and physicist language, ends with a
result. Operationalists say that a light flashes at the end of a
prepare-and-measure scenario, QBists say that an action elicits a
direct personal experience, Bohrians talk of irreversible
amplification, and even the devotees of the unitarily-evolving
All-Vector try to understand how something that looks like
irreversible amplification can emerge from the smooth cosmic
rotation. A measurement outcome is a fact, maybe a personalized or
relational one, or maybe a shared fact-for-all. How much the
generation of such a fact changes a system's physical condition is a
nontrivial question. It is worth speculating if the MGM result, seen
from an alternate perspective, may provide insight into this
question. Why not assume that the state-update rule is linear --- say,
by holding steadfast to an epistemic interpretation of quantum states
--- and then \emph{deduce} the bounds within which agnosticism can hold?

\end{document}